\documentclass[12pt]{article}
\usepackage[]{amsmath,amssymb,amsfonts,latexsym,amsthm,enumerate,fullpage}
\newtheorem{thm}{Theorem}
\newtheorem*{thm*}{Theorem}
\newtheorem{lemma}[thm]{Lemma}
\newtheorem*{lemma*}{Lemma}

\newtheorem*{prop*}{Proposition}
\newtheorem{cor}[thm]{Corollary}

\newtheorem{claim}[thm]{Claim}
\theoremstyle{remark}
\newtheorem*{rmk}{Remark}
\newtheorem*{rmk*}{Remark}
\newtheorem*{rmks*}{Remarks}
\newtheorem*{not*}{Notation}
\newtheorem*{claim*}{Claim}
\newtheorem*{fact*}{Fact}

\theoremstyle{definition}
\newtheorem{dfn}{Definition}

\def\N{\mathbb{N}}

\def\F{\mathbb{F}}
\def\P{\Pr}
\def\D{\mathcal{D}}
\def\T{\mathcal{T}}
\def\e{\mathcal{E}}

\def\x{\mathbf{x}}

\def\relwt{\texttt{rel-wt}}
\def\wt{\texttt{wt}}

\newcommand\ignore[1]{}

\begin{document}

\title{The density of weights of Generalized Reed--Muller codes}
\author{
Shachar Lovett
\thanks{Research supported by
the Israel Science Foundation (grant 1300/05).}\\
Weizmann Institute of Science\\
shachar.lovett@weizmann.ac.il }

\maketitle

\begin{abstract}
We study the density of the weights of Generalized Reed--Muller codes. Let $RM_p(r,m)$ denote the code of multivariate polynomials over $\F_p$ in $m$ variables of total degree at most $r$. We consider the case of fixed degree $r$, when we let the number of variables $m$ tend to infinity. We prove that the set of relative weights of 
codewords is quite sparse: for every $\alpha \in [0,1]$ which is not rational of the form $\frac{\ell}{p^k}$, there exists an interval around $\alpha$ in which no relative weight exists, for any value of $m$. This line of research is to the best of our knowledge new, and complements the traditional lines of research,
which focus on the weight distribution and the divisibility properties of 
the weights.

Equivalently, we study distributions taking values in a finite field, which can be approximated by distributions coming from constant degree polynomials, where we do not bound the number of variables. We give a complete characterization of all such distributions.
\end{abstract}

\section{Introduction}\label{sec:intro}
We study the weights of Generalized Reed--Muller codes.

\begin{dfn}[Generalized Reed--Muller codes]
Let $\F_p$ be a prime finite field. We denote by $RM_p(r,m)$ the
$r^{th}$-order Generalized Reed--Muller code with $m$ variables.
This is a linear code over $\F_p$, whose codewords $f \in
RM_p(r,m):\F_p^m \to \F_p$ are evaluations of polynomials over
$\F_p$ in $m$ variables of total degree at most $r$.
\end{dfn}

\begin{dfn}[Weights]
Let $\mathcal{C}$ be a code. The weight of a codeword $f \in
\mathcal{C}$ is the number of non-zero elements in it. For
$\mathcal{C} = RM_p(r,m)$, this is
$$
\wt(f) = |\{\x \in \F_p^m: f(x) \ne 0\}|
$$
\end{dfn}

One of the main problems in coding theory is understanding the
possible weights and the distribution of the weights for various families of codes.
Generalized Reed--Muller codes are one of the more basic family of codes, and
has been researched extensively. To quote \cite{MS77}:\\

{\em Reed--Muller (or RM) codes are one of the oldest and best understood families of codes}\\

Understanding the weights of codewords of Generalized Reed--Muller codes is considered
to be one of the important questions in coding theory, however our current understanding of it
 is quite limited. There are two traditional lines of research regarding the weights of Generalized
Reed--Muller codes: their distribution, and their divisibility properties. We introduce in this
work a third line of research, studying the density of the weights.

We study the weights of codewords of $RM_p(r,m)$ when we fix the order $r$ and let the number of
variables $m$ tend to infinity. This can be better described in terms of the {\rm relative weights}
of the codewords.

\begin{dfn}[Relative weights]
Let $\mathcal{C}$ be a code. The relative weight of a codeword $f \in
\mathcal{C}$ is the {\em fraction} of non-zero elements in it. For
$\mathcal{C} = RM_p(r,m)$, this is
$$
\relwt(f) = \frac{1}{p^m}|\{\x \in \F_p^m: f(x) \ne 0\}| = \P_{x \in \F_p^m}[f(x) \ne 0]
$$
\end{dfn}

Let $W_p(r,m)$ be the set of relative weights of codewords of $RM_p(r,m)$:
$$
W_p(r,m) = \{\relwt(f): f \in RM_p(r,m)\}
$$

Since $RM_p(r,m)$ can be embedded in $RM_p(r,m+1)$, we have $W_p(r,m) \subseteq W_p(r,m+1)$.
Thus it makes sense to consider the limit of the weights when $m \to \infty$. We define $W_p(r)$ to be
the set of weights of codewords of $RM_p(r,m)$ where we do not restrict $m$, i.e.
$$
W_p(r) = \bigcup_{m \in \N} W_p(r,m)
$$

The set $W_p(r)$ is contained in the interval $[0,1]$, and in fact can be further restricted based on the
minimal relative weight of $RM_p(r,m)$, which is well known (see for example~\cite{dgm70}).
We are interested however in the density of the weight set. Our a-priory intuition was that
the set $W_p(r)$ should be relatively dense, since we allow the number of variables to grow indefinitely.
However, our main theorem shows that the truth is quite far from it.  In order to state it, we first define the notion of $p$-rational numbers.

\begin{dfn}[$p$-rational numbers]
We say a number $\alpha \in [0,1]$ is {\em $p$-rational} if it is rational of the form $\alpha = \frac{\ell}{p^k}$
for some integers $\ell,k$.
\end{dfn}

\begin{thm}[Main theorem]\label{thm:main}
Let $\alpha \in [0,1]$ be a number which is not $p$-rational. Then there exists $\epsilon>0$
such that $W_p(r)$ contains no value in the interval $(\alpha - \epsilon, \alpha + \epsilon)$.
Equivalently, there is no sequence of multivariate polynomials $f_1,f_2,\dots$ over $\F_p$ of degree at most $r$,
each possibly on a different number of variables, such that $\lim_{k \to \infty} \relwt(f_k) = \alpha$.
\end{thm}

Thus, around every $\alpha \in [0,1]$ which is not $p$-rational, there is a "hole", in which there are no relative
weights of $RM_p(r,m)$.

Another way to view Theorem~\ref{thm:main} is as a theorem about the approximation of random variables over finite fields by low-degree polynomials.
\begin{dfn}[Distribution of a polynomial]
The distribution of a polynomial $f(x_1,\ldots,x_m)$ over $\F_p$ is defined to be the distribution of $f$ applied to a uniform input in $\F_p^m$.
\end{dfn}

Let $X$ be a random variable taking values in $\F_p$. We say $X$ can be approximated by degree-$r$ polynomials, if its distribution can be arbitrarily approximated by the
distribution of degree-$r$ polynomials. That is to say, for every $\epsilon>0$, there exists a multi-variate polynomial $f(x_1,\ldots,x_m)$ over $\F_p$ of total degree at most $r$, whose distribution is $\epsilon$-close to the distribution of $X$ (for example
in statistical distance). The following is an immediate corollary of Theorem~\ref{thm:main}.

\begin{cor}\label{cor:limit_by_single}
Let $X$ be a random variable taking values in $\F_p$, which can be approximated by degree-$r$ polynomials, for some constant $r$. Then all the probabilities $\P[X=a]$ are $p$-rational. In particular, $X$ can be realized as the
distribution of a single polynomial over $\F_p$.
\end{cor}

So for example, we cannot have an arbitrary good approximation of perfect random bits by constant degree polynomials over $\F_3$, for any constant degree, since $1/2$ is not $3$-rational.

Returning to the framework of weights of Generalized Reed--Muller codes, we note that although the set $W_p(r)$
is sparse, it is not finite. For example, consider the set $W_2(2)$, the set of relative weights of quadratics over $\F_2$. The relative weight of $f(x_1,\dots,x_{2k}) = x_1 x_2 + x_3 x_4 + \dots + x_{2k-1} x_{2k}$ is $\frac{2^k + 1}{2^{k+1}}$, and the set of these weights is infinite.

\subsection{Related work}
As we mentioned before, the two traditional lines of research regarding
the weights of Generalized Reed--Muller codes are studying their weight distribution and their divisibility properties. We now describe them in more details.

The weight distribution of $RM_p(r,m)$ is the number of codewords
below a certain weight. The case of $r=1$, i.e. of linear
functions, is trivial, since all non-constant codewords have the
same weight. The case of $r=2$, i.e. of quadratic functions, is
also fully understood. A theorem of Dixon~\cite{MS77} gives a
canonical characterization of quadratic functions, and in
particular gives the possible weights and the weight distribution
of quadratic functions. By the McWilliams identity, this characterize
the weight distribution of their dual codes, which are $RM_p(m-2,m)$ and
$RM_p(m-3,m)$. These are, to the best of our knowledge, the only (non-trivial) orders for which
complete characterization the weights of Generalized Reed--Muller codes is known.
For other orders, complete characterization is known only for specific values of $m$.
For example, for cubics the record is the work of Sugita, Kasami and Fujiwara~\cite{skf96}, characterizing
the weight distribution for $RM_2(3,9)$.

Considering general orders, several characteristics of the weights are known.
The minimal weight of non-zero codewords in $RM_p(r,m)$ is known,
as are as are the codewords achieving this minimal distance~\cite{dgm70}.
In the case of Reed--Muller codes, corresponding to $p=2$, Kasami and Tokura \cite{kt70} give a
complete characterization of codewords of weight at most {\em
twice} the minimal weight of the code, and Azumi, Kasami and
Tokura~\cite{akt76} gave a characterization of codewords of weight
at most $2.5$ the minimal weight of the code. Recently, Kaufman
and the author~\cite{kl08b} gave a relatively tight estimate on
the number of codewords in Reed--Muller codes, holding for all
weights.

The second line of research is divisibility of the weights of
codewords. Ax~\cite{ax64} proved that all weights of codewords $f \in RM_p(r,m)$ are
divisible by $p^{\lceil m/r \rceil - 1}$. This was later generalized to general codes~\cite{m72, dm76}. For a survey on divisible codes see ~\cite{w04} or~\cite{liu06}.

\subsection{Organization}
The paper is organized as follows. Theorem~\ref{thm:main} is proved in Section~\ref{sec:proof_main_theorem}.
The proof is based on a technical lemma which is proved in Section~\ref{sec:proof_main_lemma}.

\section{Proof of Theorem~\ref{thm:main}}\label{sec:proof_main_theorem}
We study codewords $f \in RM_p(r,m)$. Equivalently, we study polynomials:
$f$ is a polynomial over $\F_p$ in $m$ variables of total degree at most $r$.
First, we fix some notations.  We denote probabilities according to a distribution $D$ by $\P_{z \sim D}$. For a set $S$ we denote by $U_S$ the uniform distribution over $S$, and we shorthand $\P_{z \in S}$ for $\P_{z \sim U_S}$. We let $\N=\{1,2,\dots\}$ denote the set of natural numbers.
We will denote elements of $\F_p^m$ by $\x=(x_1,\dots,x_m)$, and polynomials or functions by $f(\x)=f(x_1,\dots,x_m)$.
When we refer to the degree of a polynomial, we will always mean its total degree.
The relative weight of a polynomial/function $f:\F_p^m \to \F_p$ is the fraction of non-zero elements in it,
$$
\relwt(f) = \P_{\x \in \F_p^m}[f(x) \ne 0]
$$

In order to prove Theorem~\ref{thm:main} we will show that for any degree-$r$ polynomial $f(x_1,...,x_m)$,
there exists a function $g(x_1,...,x_c)$ on a constant number of inputs (i.e. independent of $m$), such that
$\relwt(f) \approx \relwt(g)$. This is straight-forward if the required approximation is fixed a-priory; we show
this can be achieved even if the error is allowed to depend arbitrarily on the number of inputs $c$.

\begin{lemma}\label{lem:main}
Let $\e:\N \to (0,1)$ be an arbitrary mapping from the naturals to $(0,1)$. For any constant degree $r$
there exists a constant $C=C(\F_p, r, \e(\cdot))$ such that the following holds: for any degree-$r$
polynomial $f(\x)=f(x_1,...,x_m)$, there exists $c \le C$ and a function $g(x_1,...,x_c)$, such that
$$
|\relwt(f) - \relwt(g)| < \e(c)
$$
\end{lemma}

\begin{rmk}
In fact, a somewhat stronger version of the lemma also holds. Not only $|\relwt(f)-\relwt(g)| < \e(c)$.
but the statistical distance between the distributions of $f$ and $g$ is bounded by $\e(c)$.
However, we will not need this stronger version in the proof of Theorem~\ref{thm:main}.
\end{rmk}

We now prove Theorem~\ref{thm:main} using Lemma~\ref{lem:main}.

\begin{proof}[Proof of Thereom~\ref{thm:main}]
Let $\alpha \in (0,1)$ be a number which is not $p$-rational, and assume by contradiction there
exists a sequence of polynomials $f_1,f_2,\dots$ of degree at most $r$, where $f_k=f_k(x_1,...,x_{m_k})$,
whose relative weights converge to $\alpha$,
$$
\lim_{k \to \infty} \relwt(f_k) = \alpha.
$$

We now define a mapping $\delta$ from the naturals to $(0,1)$.
For every $c \in \N$, define $\delta(c)$ to be the distance of $\alpha$ from the set of rational numbers of the
form $\frac{\ell}{p^c}$. Explicitly, $\delta(c)$ is given by
$$
\delta(c) = \min \left\{ \alpha - \frac{\lfloor \alpha p^c \rfloor}{p^c}, \frac{\lceil \alpha p^c \rceil}{p^c} - \alpha \right\}
$$
Notice that $\delta(\cdot)$ is non-increasing, and by our assumption that $\alpha$ is not $p$-rational,
$\delta(c) > 0$ for all $c \in \N$.

Set $\e(c) = \frac{\delta(c)}{4}$. Once we fix the mapping $\e(\cdot)$, we can use Lemma~\ref{lem:main}:
there exists a constant $C = C(\F_p, r, \e(\cdot))$, such that
for any polynomial $f_k$ there exists $c_k \le C$, and a function $g_k(x_1,...,x_{c_k})$, such that

\begin{equation}\label{eq:A1}
|\relwt(f_k) - \relwt(g_k)| < \e(c_k) = \frac{\delta(c_k)}{4}
\end{equation}

Since $\lim_{k \to \infty} \relwt(f_k) = \alpha$, and $\e(\cdot)$ is positive, there exists some $k$ such that
\begin{equation}\label{eq:A2}
|\relwt(f_k) - \alpha| < \e(C) = \frac{\delta(C)}{4}
\end{equation}

Combining (\ref{eq:A1}) and (\ref{eq:A2}), and since $\delta(\cdot)$ is non-increasing, we get that
\begin{equation}\label{eq:A3}
|\relwt(g_k) - \alpha| < \frac{\delta(c_k)}{4}+\frac{\delta(C)}{4} \le \frac{\delta(c_k)}{2}
\end{equation}

We now show this cannot hold. $g_k$ is a function on $c_k$ inputs;

thus, its relative weight is rational of the form
$\frac{\ell}{p^{c_k}}$. By definition of $\delta(\cdot)$:
\begin{equation}\label{eq:A4}
|\relwt(g_k) - \alpha| = |\frac{\ell}{p^{c_k}} - \alpha| \ge \delta(c_k)
\end{equation}

Combining $(\ref{eq:A3})$ and $(\ref{eq:A4})$ yields a contradiction. Thus, $\alpha$ must be $p$-rational.

\end{proof}

\section{Proof of Lemma~\ref{lem:main}}\label{sec:proof_main_lemma}

The proof of Lemma~\ref{lem:main} is based on regularity results for constant degree polynomials
by Green and Tao~\cite{gt07} and by Kaufman and Lovett~\cite{kl08}. We first make some definitions.
In this section, all polynomials will be polynomials over $\F_p$ in $m$ variables.

\begin{dfn}[rank of polynomials]
Let $f(\x)$ be a degree-$r$ polynomial. The {\em $(r-1)$-rank} of $f$, denoted by $rank_{r-1}(f)$, is the
minimal number of degree-$(r-1)$ polynomials required to compute $f$. This means, $rank_{r-1}(f)$ is the
minimal $c$ such that there exists polynomials $g_1(\x),...,g_c(\x)$ of degree at most $r-1$
and a function $F:\F_p^c \to \F_p$ such that
$$
f(\x) = F\left(g_1(\x),...,g_c(\x)\right)
$$
\end{dfn}

\begin{dfn}[regularity of polynomials]
A degree-$r$ polynomial $f(\x)$ is $T$-regular if $rank_{r-1}(f) > T$. A set of polynomials $\{f_1(\x),...,f_c(\x)\}$
is $T$-regular if all non-zero linear combinations of them are $T$-regular. This means, for every $a_1,...,a_c \in \F_p$
not all zero, let $f'(\x) = a_1 f_1(\x) + \dots + a_c f_c(\x)$. We require that $f'$ is not identically zero,
and that if $degree(f')=k$, then $rank_{k-1}(f') > T$.
\end{dfn}

We will need the following result from \cite{gt07}: any degree-$r$ polynomial $f$ is a function of a constant
number of regular polynomials $g_1,\dots,g_c$, even if the regularity requirements on $g_1,\dots,g_c$
depend on the number of polynomials $c$:

\begin{lemma}[Lemma 2.3 in \cite{gt07}]\label{lem:regularity}
Let $\T:\N \to \N$ by an arbitrary mapping. There exists a constant $C_1=C_1(\F_p, r, \T(\cdot))$ such that
the following holds. For any degree-$r$ polynomial $f(\x)$ there exists some $c \le C_1$, a set of polynomials
$g_1(\x),...,g_c(\x)$ of degree at most $r$ and a function $F:\F_p^c \to \F_p$, such that:
\begin{enumerate}
  \item $f(\x) = F(g_1(\x),...,g_c(\x))$,
  \item The set of polynomials $\{g_1(\x),...,g_c(\x)\}$ is $\T(c)$-regular.
\end{enumerate}
\end{lemma}

We also need a result relating regularity of polynomials to their joint distribution.

\begin{dfn}[distribution of polynomials]
Let $f:\F_p^m \to \F_p$ be a polynomial. Its distribution $\D(f)$ is the distribution (taking values in $\F_p$) of applying
$f$ on a random input $\x \in \F_p^m$,
$$
\D(f) = f(\x)_{\x \sim U_{\F_p^m}}.
$$
For a set of polynomials $f_1,\dots,f_c:\F_p^m \to \F_p$, their joint distribution $\D(f_1,\dots,f_c)$ (taking
values in $\F_p^c$) is the distribution of applying $f_1,\dots,f_c$ on a common random input $\x \in \F_p^m$,
$$
\D(f_1,\dots,f_c) = (f_1(\x),\dots,f_c(\x))_{\x \sim U_{\F_p^m}}.
$$
\end{dfn}

\begin{dfn}[statistical distance]
Let $D',D''$ be two distributions taking values in the same set $S$. Their statistical distance is
$$
dist(D',D'') = \frac{1}{2} \sum_{s \in S}\left|\P[D'=s] - \P[D''=s]\right|.
$$
\end{dfn}

The following result from \cite{kl08} shows that polynomials whose distribution is not close to uniform
must have low rank:

\begin{lemma}[Theorem 4 in \cite{kl08}]\label{lem:bias_implies_low_rank}
Let $f(\x)$ be a degree-$r$ polynomial such that $dist(\D(f), U_{\F_p}) \ge \epsilon$. Then
$rank_{r-1}(f) \le C_2(\F_p, r, \epsilon)$.
\end{lemma}

We combine Lemma~\ref{lem:regularity} and Lemma~\ref{lem:bias_implies_low_rank} to prove the following lemma,
showing that any degree-$r$ polynomial is a function of a constant number of polynomials which are
uncorrelated.

\begin{lemma}\label{lem:comp_by_regular}
Let $\e:\N \to (0,1)$ be an arbitrary mapping from the naturals to $(0,1)$. For any constant degree $r$
there exists a constant $C=C(\F_p, r, \e(\cdot))$ such that the following holds:
For any degree-$r$ polynomial $f(\x)$ there exists some $c \le C$, a set of polynomials
$g_1(\x),\dots,g_c(\x)$ of degree at most $r$ and a function $F:\F_p^c \to \F_p$, such that:
\begin{enumerate}
  \item $f(\x) = F(g_1(\x),\dots,g_c(\x))$,
  \item $dist(\D(g_1,\dots,g_c),U_{\F_p^c}) < \e(c)$.
\end{enumerate}
\end{lemma}

\begin{proof}
We will choose $\T:\N \to \N$ large enough, to be specified later, and apply Lemma~\ref{lem:regularity}.
Let $g_1,\dots,g_c$ be the polynomials given by the lemma such that $f(\x) = F(g_1(\x),\dots,g_c(\x))$,
and the set $\{g_1,\dots,g_c\}$ is $\T(c)$-regular. We will show that if we choose $\T(\cdot)$ large enough,
we can guarantee that $\D(g_1,\dots,g_c)$ is close to uniform.

We first reduce the task to guaranteeing that all the non-zero linear combinations of $g_1,\dots,g_c$
are close to uniform.
We claim that in order to guarantee that $dist(\D(g_1,\dots,g_c) ,U_{\F_p^c}) < \e(c)$,
it is enough to guarantee for every non-zero linear combination
$g'(\x) = a_1 g_1(\x) + \dots + a_c g_c(\x)$ that $dist(\D(g'),U_{\F_p}) < p^{-c} \e(c)$.
The proof is by simple Fourier analysis: see for example Claim 33 in \cite{bv07}.

Given this reduction, we show it is enough to require that $g'$ is regular.
Assume $dist(\D(g'),U_{\F_p}) \ge p^{-c} \e(c)$. Either $g' \equiv 0$, or, by Lemma~\ref{lem:bias_implies_low_rank},if $degree(g')=k$ then
\begin{equation}\label{eq:B1}
rank_{k-1}(g') \le C_2(\F_p, k, p^{-c} \e(c))
\end{equation}

In any case, if we set $\T(c) = \max_{1 \le k \le r} C_2(\F_p, k, p^{-c} \e(c))$, we get that
the set $\{g_1,\dots,g_c\}$ is not $\T(c)$-regular, since $g'$ is not $\T(c)$-regular. This is a contradiction to the promise of Lemma~\ref{lem:regularity}.

Hence we conclude that the joint distribution $\D(g_1,\dots,g_c)$ has statistical distance of at most $\e(c)$
to the uniform distribution $\F_p^c$, where $c \le C$ and
$$
C = C_1(\F_p, d, \T(\cdot))
$$
\end{proof}

Before proving Lemma~\ref{lem:main}, we will also need the following simple claim: the statistical distance between distributions bounds
the probability that a function will be able to distinguish between them:
\begin{claim}\label{claim:statdist_func}
Let $D',D''$ be two distributions taking values in the same set $S$. Then for any subset
$S' \subseteq S$:
$$
|\P_{z \sim D'}[z \in S'] - \P_{z \sim D''}[z \in S']| \le dist(D',D'')
$$

\end{claim}

We are now ready to prove Lemma~\ref{lem:main}.

\begin{proof}[Proof of Lemma~\ref{lem:main}]
Let $f(\x)$ be a degree-$r$ polynomial.
Apply Lemma~\ref{lem:comp_by_regular}. There exists some $C=C(\F_p, r, \e(\cdot))$ such that there is
$c \le C$, a set of polynomials $g_1(\x),...,g_c(\x)$ and a function $F:\F_p^c \to \F_p$ such that
\begin{enumerate}
  \item $f(\x) = F(g_1(\x),\dots,g_c(\x))$,
  \item $dist(\D(g_1,\dots,g_c), U_{\F_p^c}) < \e(c)$.
\end{enumerate}

We claim that the function $F(y_1,\dots,y_c)$, where $y_1,\dots,y_c \in \F_p$ are independent variables,
have approximately the same relative weight as that of $f(\x)=F(g_1(\x),\dots,g_c(\x))$. We bound:

\begin{align*}
& |\relwt(f) - \relwt(F)| = \\
& |\P_{\x \in \F_p^m}[F(g_1(\x),\dots,g_c(\x)) \ne 0] - \P_{y_1,\dots,y_c \in \F_p}[F(y_1,\dots,y_c)] \ne 0| = \\
& |\P_{\x \in \F_p^m}[(g_1(\x),\dots,g_c(\x)) \in F^{-1}(\F_p \setminus \{0\})] - \P_{y_1,\dots,y_c \in \F_p}[(y_1,\dots,y_c) \in F^{-1}(\F_p \setminus \{0\})| \le \\
& dist(\D(g_1,\dots,g_c), \D(y_1,\dots,y_c)) = \\
&dist(\D(g_1,\dots,g_c), U_{\F_p^c}) < \e(c).
\end{align*}

\end{proof}

\section{Open problems}
We studied in this work the density of the weights of $RM_p(r,m)$ where we keep $r$ constant. We proved that any $\alpha \in [0,1]$ which is not $p$-rational, cannot be the limit of relative weights of constant degree polynomials. However, we can ask what is the asymptotics of the degrees of polynomials that are required to approximate $\alpha$, i.e, for every $\epsilon>0$, what should be the the degree of $f(\x)$ such that $|\relwt(f) - \alpha| < \epsilon$, and
how do this degree depend on $\epsilon$?

Another open problem is giving good bounds on the constant $C$ in Lemma~\ref{lem:main}. We note that the current proof depends on Lemma~\ref{lem:regularity} and Lemma~\ref{lem:bias_implies_low_rank}, for which no good bounds are currently known.

{\em Acknowledgements} I thank Amir Shpilka for raising the
problem studied in this paper, in the context of pseudorandom
generators for polynomials. I thank my instructor, Omer Reingold, for his constant
support and encouragement. I thank Alex Samorodnitsky, Tali Kaufman and Simon Litsyn on helpful discussions.

\end{document}